\documentclass[manuscript]{aastex631}

\shorttitle{SYMPATHETIC FILAMENTS ERUPTION AND ASSOCIATED JET}
\shortauthors{Yang et al.}

\graphicspath{{./}{figures/}}

\begin{document}

\title{Sympathetic Eruption of Two Filaments and Associated Solar Coronal Jet}

\author[0000-0003-3462-4340]{Jiayan Yang}
\affiliation{Yunnan Observatory, Chinese Academy of Sciences, P.O. Box 110, Kunming 650011, People's Republic of China}
\affiliation{State Key Laboratory of Solar Activity and Space Weather, National Astronomical Observatories, Chinese Academy 
of Sciences, Beijing 100101, People's Republic of China; 
lepingli@nao.cas.cn}

\author{Leping Li}
\affiliation{State Key Laboratory of Solar Activity and Space Weather, National Astronomical Observatories, Chinese Academy 
of Sciences, Beijing 100101, People's Republic of China;
lepingli@nao.cas.cn}

\author{Huadong Chen}
\affiliation{State Key Laboratory of Solar Activity and Space Weather, National Astronomical Observatories, Chinese Academy 
of Sciences, Beijing 100101, People's Republic of China;
lepingli@nao.cas.cn}
\affiliation{State Key Laboratory of Solar Activity and Space Weather, National Space Science Center, Chinese Academy 
of Sciences, Beijing 100190, People's Republic of China}

\author{Yi Bi}
\affiliation{Yunnan Observatory, Chinese Academy of Sciences, P.O. Box 110, Kunming 650011, People's Republic of China}
\affiliation{Yunnan Key Laboratory of Solar Physics and Space Science, Kunming 650011, People's Republic of China}

\author{Bo Yang}
\affiliation{Yunnan Observatory, Chinese Academy of Sciences, P.O. Box 110, Kunming 650011, People's Republic of China}
\affiliation{Yunnan Key Laboratory of Solar Physics and Space Science, Kunming 650011, People's Republic of China}

\author{Junchao Hong}
\affiliation{Yunnan Observatory, Chinese Academy of Sciences, P.O. Box 110, Kunming 650011, People's Republic of China}

\author{Yan Dong}
\affiliation{Yunnan Observatory, Chinese Academy of Sciences, P.O. Box 110, Kunming 650011, People's Republic of China}

\begin{abstract}

Combining the high-quality observations from the {\it Solar Dynamics Observatory} (SDO), the Global Oscillation Network 
Group (GONG), and the Chinese H$\alpha$ Solar Explorer (CHASE), we report a solar coronal jet triggered by the sympathetic 
eruption of two filaments on 2024 January 11. Initially, the western segment of an active region filament erupted. The 
erupting plasma propagated eastward, approximately along the filament's axis. This eruption perturbed the magnetic field of 
a second filament situated near its eastern footpoint, the second filament then erupted sympathetically about one hour 
later. The eruption of the second filament is a failed one, with the majority of the filament material falling back 
after the initial lifting. Although no GOES flare accompanied these filament eruptions, distinct brightenings were observed 
following each eruption. The second eruption produced a large coronal jet, which propagated along a bent trajectory with 
an apparent deflection angle of approximately 90 degrees. No clear evidence of magnetic reconnection was detected at 
the deflection site, thus we suspect that the jet may have traveled along an S-shaped trans-equatorial loop and shown 
a curved trajectory. This event exhibits multiple phenomena: partial filament eruption, failed filament eruption, sympathetic 
filament eruption, jet initiation by filament eruption, and apparently deflected jet propagation. Collectively, these 
observations highlight the complexity and diversity of solar activity.

\end{abstract}

\keywords{Solar activity (1475)  --- Solar filament eruptions (1981) --- Solar magnetic fields (1503)
--- Solar radiation (1521)}

\section{Introduction} \label{sec:intro}
Solar filaments/prominences are ubiquitous structures in the solar atmosphere. They are long, slender sheets,
rooted in the chromosphere and suspended in the corona. The density of filaments is approximately 100 times that of the 
ambient corona, while their temperature is only about one percent of the coronal temperature \citep{labrosse10}. Therefore, 
the magnetic field topology plays a crucial role in supporting the cold, dense filament material floating in the hot, 
tenuous corona \citep{mackay10}. When the supporting magnetic field is disturbed or changed, the filament may lose 
equilibrium and further erupt in many cases.

Filament eruptions are one of the most energetic explosive phenomena in the solar atmosphere. They are often closely
associated with solar flares and coronal mass ejections (CMEs) \citep{chen25, duchlev21}, thus have significant influence 
on space weather. Filament eruptions have been studied extensively for decades, but they continue to attract substantial 
research interest, and many key problems have remained unresolved till now \citep{chen20b}. In some cases, two 
or more adjacent filaments erupt one by one within a short time interval, and some specific physical linkage exist 
between/among them. Such special filament eruptions are termed sympathetic filament eruption \citep{jiang11, yang12}, 
following the nomenclature of sympathetic flare \citep{fritzova76, mawad22} and sympathetic CME \citep{moon03, lugaz17}.
 
Sympathetic eruptions are not uncommon in the solar atmosphere. \cite{guite25} conducted a statistical study of over 
40000 flares observed by three space-borne solar telescopes during solar cycles 24 and 25, finding that approximately 
5 percent of them were sympathetic flares. They also deduced that the peak angular separation of the sympathetic flares 
in longitude was around 30\degr, with most events they studied exhibiting a time interval of not exceeding 1.5 hours
between two successive flares. Observational case studies showed that, two or more eruptions in a sympathetic event 
not only occur in close spatial and temporal proximity but also exhibit a physical linkage with each other \citep{green18}. 
In our previous studies of two sympathetic filament eruption events \citep{jiang11, yang12}, coronal dimmings appeared 
during the eruption of the first filament, extending to the second filament's location and persisting. Based on these 
observations, we proposed that diffuse coronal dimmings -- which reflect the expansion of magnetic loop systems -- could 
act as a linkage in sympathetic eruptions. We further suggested that a multiple-arcade helmet streamer is a favorable 
magnetic configuration for sympathetic eruptions. Other possible physical connections between sympathetic eruptions may 
be a set of propagating disturbances \citep{wang01}, change of the large-scale magnetic field topology caused by the
magnetic implosion mechanism \citep{shen12a, zhou21}, filament interactions and the opening or weakening of overlying 
coronal loops by the initial eruption \citep{kong13, xue14, xue16}, expansion and reconnection of the overlying 
large-scale magnetic field \citep{joshi16, wang18}, continuous intrusion of flare ribbon formed after the first 
eruption \citep{song20}, oscillation and mass drainage of the second filament \citep{dai21}, and EUV waves \citep{li25b}. 

Consistent with observations, numerical simulations confirmed that the physical linkage between sympathetic eruptions 
is fundamentally magnetic in nature. By combining stereoscopic coronal imaging observations with potential field 
source surface (PFSS) extrapolation, \cite{schrijver11} investigated more than ten sympathetic eruption events occurring 
on 2010 August 1-2. They identified multiple direct magnetic connections between the two eruptions of a sympathetic event, 
manifested as magnetic separatrices, separators, and quasi-separatrix layers. A subset of five primary eruptions from 
these events were studied in detail by \cite{titov12} using the PFSS modeling. Their simulations revealed the presence 
of several pseudo-streamers with associated magnetic null points and separators, suggesting that the magnetic reconnection
triggered by the initial eruption at these null points and separators may serve as the key mechanism linking sympathetic
eruptions. This suggestion was further supported by high-resolution 2.5-dimensional (2.5D) magnetohydrodynamic (MHD) 
simulation presented by \cite{lynch13}, which demonstrated that magnetic null points in pseudo-streamers naturally facilitate 
sympathetic eruptions. \cite{torok11} conducted a 3-dimensional (3D) MHD simulation to reproduce one of the sympathetic 
events of \cite{schrijver11}, and proposed two potential magnetic triggering mechanisms of sympathetic eruptions: magnetic 
reconnection, and the weakening of the common overlying field. Based on a larger event sample, \cite{schrijver13} further 
identified four fundamental magnetic linkage pathways governing sympathetic eruptions.
 
In addition to their well-established connection with solar flares and CMEs, recent studies have demonstrated that eruptions of 
small-scale filaments or mini-filaments are also closely associated with coronal jets \citep{hong11, hong16, shen17, yang19,
yang20, duan22}. Traditionally, solar coronal jets were thought to be primarily launched through magnetic reconnection between 
newly emerged dipolar magnetic field and pre-existing background field, under the framework of the emerging-flux model 
\citep{shibata94, yokoyama95, shen21}. Based on a statistical study of 20 coronal jets, \cite{sterling15} provided
clear evidence that solar coronal jets could also be triggered by the small-scale filament eruptions in source 
regions. This alternative mechanism was subsequently supported by 3D MHD simulational studies of \cite{wyper17, wyper18}. 
It is now widely acknowledged that magnetic flux emergence and small-scale filament eruptions are the two fundamental
mechanisms to trigger solar jets. \cite{wyper17} further proposed that the physical mechanisms of CMEs produced by 
large-scale filament eruptions and jets produced by small-scale filament eruptions are essentially identical. Moreover,
observations reveal that while most jets either fall back to their source regions or propagate along closed loops to other
locations at solar surface after eruptions, there are still some jets that can escape the solar surface and be detected  
as narrow CMEs in white-light coronagraph observations \citep{sterling18, yang20}. Particularly, in rare cases both narrow 
jet-like CME and bubble-like CME could be produced simultaneously by small-scale filament eruptions \citep{shen12b, duan19}.

On 2024 January 11, a small filament erupted in the northwest part of the solar disk, triggering the sympathetic eruption
of another filament located to the east of it. Furthermore, the second filament's eruption produced a coronal jet exhibiting
significant apparent trajectory deflection. While initially appearing as a minor event, this filament eruption ultimately 
generated a series of interconnected phenomena such as sympathetic eruption, jet initiation, and apparently deflected 
plasma propagation. Therefore, this event provides valuable insights into the complexity of solar activity. The paper is 
organized as follows: Section 2 describes the observational instruments and analysis methods, Section 3 presents the 
detailed observations and our analysis of the event, and Section 4 summarizes the findings and discusses their implications for 
understanding sympathetic eruptions and jet dynamics.

\section{Instruments and Method}\label{sec:instru}
This study primarily utilizes level 1.5 full-disk extreme-ultraviolet (EUV)/ultraviolet (UV) images from the Atmospheric 
Imaging Assembly \citep[AIA;][]{lemen12} on board the {\it Solar Dynamics Observatory} \citep[{\it SDO};][]{pesnell12}, 
combined with H$\alpha$ line-center images from multiple instruments, to analyze the sympathetic filament eruptions and 
the associated jet. AIA takes images at 7 EUV passbands and 2 UV passbands, with a pixel size of $0^{''}.6$. The time cadence 
is 12 seconds for EUV images and 24 seconds for UV images. The EUV images obtained at 304 \AA, 171 \AA, 211 \AA, and 335 \AA\, 
along with the UV images obtained at 1600 \AA\ are employed for this investigation. H$\alpha$ data used here are obtained from 
the Global Oscillation Network Group \citep[GONG;][]{harvey96} at the National Solar Observatory and the H$\alpha$ Imaging 
Spectrograph \citep[HIS;][]{liu22} on board the Chinese H$\alpha$ Solar Explorer \citep[CHASE;][]{li19, li22}. CHASE/HIS 
operates in two observational modes: raster scanning mode and continuum imaging mode. Here we utilize the full-Sun spectral 
images from H$\alpha$ waveband (6559.7 - 6565.9 \AA\ ) obtained by the raster scanning mode, and extract H$\alpha$ line-center 
images for further analysis. The pixel spatial resolution of these images is $0^{''}.52$, and the temporal resolution is 
1 minute. The time cadence of images provided by GONG is the same as that of CHASE/HIS, with the pixel size of $1^{''}$.

To investigate the magnetic environment surrounding the two filaments and subsequent jet, we analyze the full-disk 
line-of-sight (LOS) magnetograms obtained by the Helioseismic and Magnetic Imager \citep[HMI;][]{scherrer12} on board the 
{\it SDO} to characterize the photospheric magnetic fields. The pixel size is $0^{''}.5$ and the time cadence is 
45 seconds for HMI LOS magnetograms. For characterizing the large-scale coronal magnetic topology surrounding the 
filaments, we perform the Nonlinear Force-Free magnetic Field (NLFFF) extrapolation \citep{wiegelmann04, wiegelmann06} 
using HMI Active Region Patches \citep[HARPS;][]{bobra14} vector magnetograms as the photospheric boundary condition.

In addition, we also check the X-ray flare list from the Geostationary Operational Environmental Satellite 
\citep[GOES;][]{hill05} and the CME catalogs from the Large Angle and Spectrometric Coronagraphs \citep[LASCO;][]{brueckner95} 
on board the {\it Solar and Heliospheric Observatory} \citep[SOHO;][]{domingo95}. However, it seems that neither 
recorded flares nor CMEs were correlated with this event.

\section{Observational Results}\label{sec:result}
\subsection{Overview}
The event studied here occurred on 2024 January 11, when two filaments erupted sequentially on the solar disk,
triggering a conspicuous coronal jet with a significantly apparently deflected trajectory. No associated flare or CME 
was recorded. Figure 1 presents an overview of the region at 00:26 UT, about 90 minutes before the filament eruptions. As 
revealed by AIA 304 \AA, 171 \AA, and CHASE H$\alpha$ line-center images in the top row, the two filaments F1 and F2 were 
located in the northwest quadrant of the solar disk, relatively close to the Sun's equator and western limb. The CHASE H$\alpha$ 
image (panel (b1)) shows that a large curved quiescent filament QF was situated to the south of F1 and F2. The white dotted 
curves indicate the trajectory of the jet triggered several hours later by the eruptions of F1 and F2. While QF appears
to lie along the path of the jet, this alignment may result from an illusion caused by the projection effect, suggesting
the jet and QF may occupy different heights and positions of the corona. In reality, QF exhibited clear disturbances during 
the courses of the filament eruptions and the jet but ultimately remained stable. The HMI LOS magnetogram (panel (d1))
reveals that three active regions (ARs) were involved in this event: AR 13536 and AR 13539 in the northern hemisphere, 
and AR 13543 in the southern hemisphere. As shown in the images in the bottom row, zoomed-in views of the region marked 
by white rectangles in the top row provided clearer visualization of filaments F1 and F2. It is noticed that in AIA 171 
\AA\ image (panel (c2)), the two filaments appeared as a single structure, whereas in 304 \AA\ image (panel (a2))
they are clearly distinct. This separation became unequivocal when they began to lift and erupt subsequently (Figure 2), 
confirming F1 and F2 were two filaments rather than two sections of a single filament. Their projected lengths 
measure approximately $4.72 \times 10^{4}$ km for F1 and $2.81 \times 10^{4}$ km for F2, respectively. Morphologically, 
F1 exhibits a winding appearance compared to F2's smoother arched shape, likely corresponding to greater magnetic shear 
in F1's underlying photospheric magnetic field (panel (d2)). When overlaying filament axes obtained from AIA 304 \AA\ 
image (panel (a2)) by manual visual inspection on the simultaneous HMI magnetogram (panel (d2)), it is found that F1 
lies above the main neutral line of AR 13536, while F2 is positioned along the polarity inversion line between the positive 
polarity of AR 13539 and the negative polarity of AR 13536. Therefore, F1 qualifies as an AR filament whereas F2 may 
represent an intermediate filament \citep{yan15}. Both the EUV images overlaid with magnetic contours and the magnetogram 
overlaid with the outlines of filaments' axis demonstrate that, the eastern footpoint of F1 anchors in negative polarity 
region and its western footpoint roots in positive polarity region, while F2's southeastern footpoint associates with 
negative polarities and its northwestern footpoint links to positive polarities.

\subsection{Sympathetic eruption of two filaments}
Following 02:00 UT, filaments F1 and F2 erupted successively within approximately one hour, with no associated GOES
flares or CMEs recorded. The close temporal and spatial connections between these two eruptions, along with clear 
physical linkage evidence presented subsequently, strongly suggest a sympathetic filament eruption scenario.
The detailed process of the sympathetic filament eruptions is displayed in Figure 2, where panels (a1-a8) focus on the
eruption of F1 while panels (b1-b7) focus on the eruption of F2. It is revealed that F1's activity began with a localized 
brightening near its western end around 01:52 UT (panel (a1)) in AIA 304 \AA\ observations. Superimposing the 
outlines of the filaments' axes obtained from AIA 304 \AA\ image at 00:26:53 UT by manual visual inspection onto 
this panel, it can be seen that the shapes and the locations of both F1 and F2 are basically unchanged yet. The western 
portion of F1 then began to lift and erupt, and the brightening spread and became brighter. However, the main filament 
body remained stable even in the subsequent process, thus F1 experienced a partial eruption \citep{chen18}. AIA 171 \AA\ 
images (panels (a2) and (a6)) revealed numerous coronal loops existed in the region, including a group of loops 
overlying F2 and a bright loop (BL) anchored west of F1. These loops played a role in the eruption dynamics of 
F1 and the subsequent jet formation.

The erupting plasma initially propagated westward, approaching BL by approximately 02:12 UT (panel (a6)), potentially
colliding with it before subsequently redirecting eastward (see panels (a7-a8) and the accompanying animation). 
While the brightening surrounding F1's west end gradually diminished, a weak eastward plasma flow continued approximately 
along F1's axis. Typically, at 03:02:17 UT, it is noticed that a distinct bright filamentary feature (Br1) reached F2's 
vicinity (panel (b1)), and passed beneath it subsequently (panel (b2)). Within two minutes, some bright patterns appeared
around the northwestern footpoint of F2 and gradually spread to the other end, triggering F2's eruption. The spatial 
proximity between F2's northwestern footpoint and F1's eastern footpoint suggests that, the sustained eastward plasma
flow following F1's eruption perturbed the local environment near the northwestern footpoint of F2, causing the
footpoint to rise and ultimately destabilizing the entire F2, leading to its eruption. In this sense, the 
eruptions of F1 and F2 are sympathetic, with inter-filament plasma transport serving as the primary physical linkage. 
Notably, a loop-like brightening (Br2) formed ahead of the erupting F2 (see panel (b5) for instance, but this brightening 
is discernible across all AIA EUV passbands). Given F2's envelopment by dense coronal loops, the occurrence of Br2 likely 
marks the interaction between the erupting F2 and overlying magnetic structures \citep{li25a}, and this interaction is 
potentially critical for initiation of the following jet. The subsequent 304 \AA\ images showed that F2 initially erupted 
northwestward, but then fell back. Therefore, different from the partial eruption of F1, F2 experienced a failed eruption 
\citep{ji03, yan20}. Accordingly, by 04:00 UT (panel (c)) both F1 and F2 remained visible in the region: F1 maintained 
approximately its original position, while F2 settled slightly displaced from its pre-eruption location.

To better characterize the temporal and spatial relationship between the two filament eruptions, we constructed a series
of time-space plots along a curved slit AB (marked by the blue dotted curve in panel (a3) of Figure 2) using AIA 304, 171, 
211, and 335 \AA\ images. These time-slice plots are presented in Figure 3. The eruption trajectory of F1 (stressed by the 
blue dotted broken line in panel (b)) is similar in all panels, and it reveals that F1 initially ejected towards point B 
before turning toward point A (thus toward the location of F2), with the cold filament plasma mainly ejected toward 
point A following the hot plasma. The plots also clearly capture the rise and fall phases of F2, which align perfectly 
with the imaging observations in Figure 2. It is inferred that F2's eruption onset occurred at about 03:07:12 UT, as 
indicated by the vertical white dashed line in panel (c). Furthermore, the time-slice plots provide compelling evidence 
of sustained plasma transport between the filaments, showing persistent mass flow along slit AB toward F2 that continued 
until the eruption of F2. Some episodes of these mass flows are indicated by the white arrows in panel (a) and the white 
dotted lines in panel (b). Therefore, the multi-wavelength time-slice plots confirm the imaging results, providing further 
verification of the physical linkage between the eruptions of F1 and F2.

The eruptions of both F1 and F2 were each accompanied by distinct flare activity. Although too weak to be registered in 
GOES flare records, both flares were clearly captured in SDO/AIA and GONG H$\alpha$ observations. Figure 4 presents the 
flares in AIA 1600 \AA\ and GONG H$\alpha$ line-center images. It shows that the flare associated with F1's
eruption (FL1) is a compact one, whereas the flare associated with the eruption of F2 (FL2) exhibited more structured 
ribbon patterns. Besides the flare ribbons, the curved brightening Br2, visible ahead of erupting F2 in all AIA EUV 
passbands as we mentioned above, also appears distinctly in AIA 1600 \AA\ images (marked by the white arrow in panel (a4)). 
This brightening is indiscernible in GONG H$\alpha$ data, indicating that its temperature is higher compared to the 
flare emissions. The normalized light curves of 1600 \AA\ and six EUV wavebands (excluding 193 \AA\ due to data artifacts) 
during the course of two filament eruptions are presented in panel (c). The fluxes are derived from integrated intensities 
across the entire FOV of the images shown in this figure. The light curves of different wavebands demonstrate two 
distinct peaks nearly synchronous, corresponding to the two flares FL1 and FL2 respectively. In addition, it is 
inferred that FL2 was brighter and more energetic than FL1 in all wavebands we investigated here.

\subsection{The associated jet and its apparently deflective propagation}
Figure 5 and its accompanying animation demonstrate the dynamic process of a prominent coronal jet generated by F2's 
eruption. It has been widely accepted in recent years that coronal jets can be triggered by the eruption of 
small-scale filaments, but the filament and the associated jet in this case exceeded the typical dimensions of filament-driven 
jets reported in previous studies \citep{hong11, sterling15, shen17, yang19}. The jet's complete production and evolution 
processes are shown by AIA 304 \AA\ and 171 \AA\ observations (panels (a1-b4)). We note that the jet first 
ejected at about 03:04 UT when the northern footpoint of F2 just started to rise, manifesting itself as a bright straight 
streak in AIA 304 \AA\ images (panels (a1) and (b1)). Filament F1 and the large quiescent filament QF are also visible at 
that moment. The jet plasma first streamed southeastward, approximately following the polarity inversion line between ARs 
13536 and 13539 (see panel (d1) of Figure 1). At 03:09 UT, the whole F2 lifted and the curved brightening Br2 which is 
visible across all AIA EUV channels appeared ahead of it (panel (a3)). The occurrence of Br2 may signify the magnetic 
reconnection between the magnetic field of erupting F2 and the overlying coronal loops \citep{Xu24, li25a}. It is suggested 
that magnetic reconnection is a primary trigger mechanism of solar jets, thus the occurrence of Br2 may also indicate the 
beginning of the main eruption phase of the jet. Following this reconnection signature, substantial amount of multi-thermal 
plasma ejected from the site, and the jet exhibited characteristic features of a blowout jet \citep{moore10} with widening 
spire and pronounced untwisting motions. At the same time, flare ribbons FL2 appeared around F2 with bright post-flare loops 
connecting them (panels (a4) and (a5)). The nearby quiescent filament QF experienced obvious disturbances throughout this 
process, displaying violent plasma motion along its spine and significant brightening, yet maintained its structure without 
erupting.

Subsequent observations revealed a striking apparent 90-degree deflection in the jet's trajectory, shifting from its 
initial southeastward propagation to southwestward direction. AIA 171 \AA\ running difference image show that the 
apparent deflection occurred around 03:15 UT (panel (b3)). The jet's bent morphology is apparent in both direct 
and fixed-base difference 304 \AA\ images. While the deflected propagation of CMEs in the low corona has been studied 
extensively for several decades since the 1980s \citep{macqueen86, cremades04}, the underlying mechanism for jet 
deflection may differ substantially, a point we will elaborate on in the following section. During the post-deflection phase, 
AIA 304 \AA\ observations demonstrated that the jet spire was dominated by dark, cold plasma (panels (a6)-(a8)). Therefore, 
although the eruption of F2 failed and F2 eventually fell back, significant mass from F2 was successfully channeled 
into the jet.

According to previous studies, some coronal jets can escape the solar atmosphere and be detected as CMEs \citep{shen12b,
sterling18, duan19, yang20}. However, in our case no associated CME was observed by LASCO C2 despite the jet's apparent 
extension beyond the solar limb in its late phase (see panel (a7) for instance). Considering that the jet was located 
fairly close to the western edge of the solar disk, we suspected that the jet's apparent reaching the edge may be an optical 
illusion caused by the projection effect. Fortunately, after checking several days' AIA observations before January 11, we 
found that a homologous jet occurred the previous day (2024 January 10) when the source region was positioned farther 
from the limb, and the AIA 304 \AA\ images presented in panels (c1-c3) demonstrate the general appearance of this homologous 
jet. The striking positional and morphological similarities between these two jets strongly suggest that they propagated 
along the same large-scale coronal loop system. By tracking the homologous jet's trajectory in AIA 304 \AA\ movies (not 
available in this paper), the approximate position of its southern footpoint was identified and marked by the white circles 
in the bottom row of Figure 5. When superimposed on the HMI LOS magnetogram (panel (c4)), it is inferred that the 
southern end of the jet was rooted in a weak, dispersed magnetic field region, located to the south of AR 13543. 
This configuration conclusively demonstrates that the January 11 jet followed a closed magnetic topology rather than open 
field lines, thus no CME was produced after the jet. This homologous event also advanced our understanding of how apparently 
open structures near the solar limb may actually represent closed magnetic systems when viewed in proper 3D context.

The jet exhibited complex kinematic behavior besides its apparent trajectory deflection, as revealed by the 
animation associated with Figure 5. The jet spire experienced untwisting motion as in many blowout jets \citep{hong16, 
shen21, yang19}, along with a portion of jet plasma flowing back to the source region along the original eruption path. 
To quantify these dynamics, we constructed time-slice plots using two slits S1 and S2 indicated by the long blue arrows in 
Figure 5, and presented them in Figure 6. Panels (a) and (b) show time-slices from the bent slit S1, which is 
aligned with the propagation direction of the jet. Both the ascending and descending phase of the jet plasma are discernable 
in the time-slices, as pointed out by the white short arrows. Linear fits (the white dotted lines) to the trajectories 
yield projected velocities of $349.11 \pm 9.15$ km s$^{-1}$ for the ascending phase and $91.52 \pm 2.45$ km s$^{-1}$ 
for the descending phase. These values align well with previous studies of solar EUV jets \citep{shen21}. For example, in a 
statistical study of \cite{nistico09}, the typical velocities of energetic polar EUV jets are around $400$ km s$^{-1}$ in 
171 \AA\ observations and $270$ km s$^{-1}$ in cooler 304 \AA\ observations. Similarly, \cite{mulay16}, using SDO data, 
found velocities of 20 AR jets ranging from $87$ to $532$ km s$^{-1}$. In a study of 12 recurrent jets in AR 11726, 
\cite{yang23a} measured speeds between $80.6$ km s$^{-1}$ to $433.6$ km s$^{-1}$ during ejection phases and $56.2$ km s$^{-1}$ 
to $101.9$ km s$^{-1}$ during falling phases, further supporting the consistency of our measurements with established jet 
velocities. The start time of the jet inferred from AIA 171 \AA\ time-slice (panel (b)) is 03:10:12 UT, while AIA 304 \AA\ 
observations revealed that the jet first appeared at 03:04 UT as a spiny bright feature (Figure 5). This time lag is due to 
the fact that S1's position missed the early peripheral ejection of the jet. The spire of the jet in this event is 
relatively wide, and the slit S1 we chose here passed through its central part and thus represented the movement of 
the main part of the jet but missed the initial signal of the jet. The time-slice shown in panel (c) is built along 
a straight slit S2 positioned near the jet base and approximately perpendicular to the jet spire (see panel (a8) of 
Figure 5), thus we can detect the transverse motion of the jet. It shows that at least two episodes of untwisting 
motions could be resolved, as pointed out by the short black arrows. Together, these time-slice plots reveal the jet's 
multi-component nature: rapid outward flow of hot, twisted plasma followed by slower draining movement of cold plasma along 
the same path.

\section{Conclusion and Discussion}\label{sec:discussion}
This study presented a comprehensive analysis of interconnected solar eruptions occurring on 2024 January 11. The events 
were initiated with the partial eruption of a small squiggly filament F1 located roughly along the main neutral line of 
AR 13536 in the northern hemisphere. The erupted plasma moved eastward, disturbing the magnetic environment 
at the northwest footpoint of another filament F2, leading to F2's subsequent sympathetic eruption. While each eruption 
produced localized brightenings, neither of them was bright enough to be recorded as a GOES flare. F2 exhibited 
typical failed eruption characteristics, with its main body falling back about ten minutes after eruption. 
However, a coronal jet was launched by F2's failed eruption, and a portion of cold filament plasma was ejected 
into the jet. The resulting jet displayed complex kinematics, including a curved trajectory with an apparent 
deflection angle of about 90 degrees, obvious lateral untwisting motions, as well as the back and forth motion along 
its spire. Through comparison with a homologous jet observed the previous day, we established that both jets 
propagated along the same large-scale closed coronal loop, explaining the absence of an associated CME despite the jet's 
apparent limb extension.

As proposed by previous studies, sympathetic filament eruptions are closely related to the surrounding magnetic field 
topology. Therefore, we further investigated the coronal magnetic field configuration above the filaments 
before and after the eruption of F1. Figure 7 present the extrapolated magnetic 
field lines obtained through NLFFF extrapolation \citep{wiegelmann04, wiegelmann06} applied to HMI vector magnetograms, 
overlaid on an AIA 304 \AA\ image taken at 01:44:53 UT. Panels (a1-a2) show the coronal magnetic configuration 
at 01:48 UT when both filaments were still stable, while panels (b1-b2) depict the same region at 02:36 UT, a time when F1 
had undergone partial eruption but F2 remained inactive. As shown by the grey curves in panels (a1) and (a2), a 
low-lying twisted structure is present prior to the eruption, aligning well with both the position and the observed 
morphology of filament F1 in 304 \AA\ images. After the partial eruption of F1, the western portion of the twisted 
structure exhibited a clear reduction in twist (panels (b1) and (b2)), consistent with the observed partial eruption 
behavior of F1. As mentioned above, AIA EUV and CHASE H$\alpha$ observations showed that filament F2 was located at 
the east of F1. However, we failed to deduce another twisted magnetic structure at the corresponding region, but obtained 
a magnetic arcade overlying F2 (shown by the dark blue curves) instead. Considering the weak shear of the photospheric 
magnetic field below F2 (see panel (d2) of Figure 1 for instance), it may be hard to deduce a twisted structure above this
 broad polarity inversion line by NLFFF extrapolation. 

The extrapolated coronal magnetic configuration further revealed that both F1 and F2 share a common overlying arcade, as 
indicated by the cyan curves. The role of a common overlying arcade in promoting sympathetic filament eruptions has been 
documented in previous studies \citep{kong13, xue14, xue16}. In such cases, the first eruption perturbs the common arcade, 
thereby weakening the magnetic confinement of the other filament and facilitating the second eruption. However, in
the present case, although F1 and F2 were both covered by a common overlying arcade, the extrapolated magnetic field lines
of this arcade remained nearly unchanged before and after the eruption of F1 (see panels (a1-a2) and (b1-b2)), suggesting 
that the common arcade may be not significantly perturbed. To more precisely analyze the impact of F1's eruption on the 
overlying common arcade, we further calculated the decay index above F1 and F2 along the approximate spines of them 
(markeded by the red line P1P2 in panel (a1)). The resulting height-dependent decay index distributions are presented in the 
bottom row of Figure 7 (see panels (a3) and (b3)), with the corresponding color bar shown on the right. It is well 
estabished that a flux rope becomes unstable where the decay index exceeds approximately 1.5 \citep{kliem06}, a threshold 
corresponding to the red and yellow regions in our figure. The white dotted lines in panels (a3) and (b3) mark
the height of the central portion of the extrapolated flux rope, approximately corresponding to the height of F1. It is shown
that the most of F1 lies within a region where the decay index remains below 1.5, except for a small segment indicated by 
the black line. We mapped this region onto line P1P2 in panel (a1) and highlighted it with a blue line. This region clearly 
corresponds to the eastern end of F1, which is expected to be located at a significantly lower height than the central part of the 
flux rope (marked by the white dotted lines). This implies that this segment of F1 is also situated in a magnetically stable 
region, lying below the black line shown in panels (a3) and (b3). Although the height of F2 could not be determined directly
due to the failed extrapolation of its flux rope, it lies east of F1 in a region where the decay index remains below the 
stability threshold even at altitudes significantly exceeding that of F1, suggesting that F2 was also magnetically confined. 
A comparison between panels (a3) and (b3) indicates almost no 
significant change but only several minor variations in some places in the decay index following 
F1's eruption, further confirming that the common arcade may be not substantially weakened by the partial eruption of F1. 
This result may be caused by two factors: either the extrapolated coronal magnetic field and decay index are both derived from 
the photospheric magnetic field which remained mostly unchanged after the eruption, or the coronal magnetic field itself indeed 
experienced minimal disturbance, as the eruption was relatively weak and largely confined to F1's western end. Therefore, 
there is insufficient evidence to support the common overlying arcade as a physical linkage of the sympathetic 
filament eruption in this event. Instead, the persistent disturbance around the northwestern footpoint of F2 caused by the 
eastward plasma flow from F1 appears to be the primary mechanism responsible for triggering the sympathetic eruption of F2. 
This process was clearly captured in AIA observations (Figure 2) and also demonstrated in the time-space plots (Figure 3): 
after the partial eruption of F1, multi-thermal plasma flows containing both hot and cold components traveled from F1 to F2, 
followed by localized brightenings at F2's northwestern footpoint that preceded its destabilization and eruption. 

Regarding the contribution of plasma flow from F1 to F2 in triggering F2's sympathetic eruption, we propose two potential 
mechanisms here. First, the continuous plasma flow may interact with the supporting magnetic field of F2, destabilizing the 
magnetic environment near its northwestern footpoint and promoting its eruption. Second, the sustained material injection 
could increase the density and gas pressure beneath F2, disrupting its force equilibrium and initiating its ascension. 
Our quantitative analysis yields an estimated gas pressure increase of approximately $5.0$ dyn cm$^{-2}$ beneath F2, compared 
to its magnetic pressure of about $15.92$ dyn cm$^{-2}$, a value consistent with measurements reported by \cite{chen20a} for 
an activated solar tornado-like prominence. The detailed derivation process is provided in the Appendix. However, as 
noted therein, both pressure estimates are subject to substantial uncertainties that may extend to an order of magnitude levels.  
While we cannot precisely quantify the contribution of this gas pressure enhancement to destabilizing F2, our results only 
suggest that it may serve as a perturbation factor that cooperated with other mechanisms to trigger F2's eruption.

It should be noted that sympathetic filament eruptions are different from filament interactions \citep{yang17, yang23b}. 
In short, the bodies of two filaments usually contact directly for filament interactions, while the correlation between two 
filaments in a sympathetic eruption is usually implicit and indirect \citep{song20}. The time interval of two filament 
eruptions is approximately one hour, aligning well with statistical findings that the interval for sympathetic flare
does not exceed 1.5 hours \citep{guite25}. 

The deflection of CMEs in the low corona has been extensively investigated over several decades \citep{green18, mierla23,
zhang21, zhang22}, and it has been widely accepted that the ambient coronal magnetic field configuration fundamentally 
determines CME trajectories within a few solar radii \citep{cecere20, cecere23}. Generally speaking, CMEs always propagate
towards the direction of least resistance, or rather, the region with lower magnetic energy density \citep{gui11, lynch13, 
shen11, sieyra20}. Therefore, open fields of coronal holes or strong magnetic fields from ARs would keep CMEs 
away from them \citep{cremades06, gopalswamy09, sahade20, sahade21}, while heliospheric current sheet, helmet streamers,
pseudostreamers or other weak field regions would guide CMEs to move towards them \citep{sahade22, sahade23, wang20}. 

Besides CMEs, solar jets have also been observed to eject following curved trajectories occasionally. As 
transient, collimated plasma beams constrained to follow local magnetic field lines, jets primarily trace the pre-existing 
magnetic field lines rather than being attracted or obstructed by surrounding magnetic structures like CMEs. \cite{nistico15} 
investigated the propagation of 79 polar coronal hole jets during solar minimum and revealed systematic low-latitude 
deflections with notable hemispheric asymmetries in deflection patterns. The observations are consistent with the fact that 
the polar magnetic field lines were bent toward the solar equator. However, exceptional cases do exist where jets undergo 
deflections through direct interaction with surrounding magnetic structures. \cite{zheng16} reported an extreme instance where 
a jet collided with a coronal hole was abruptly redirected in the opposite direction, while \cite{li20} captured the 
detailed deflection and splitting process of a chromospheric fan-shaped jet upon encountering nearby facular magnetic structure. 
Similarly, \cite{mitra20} observed drastic deflection of jet-like plasma ejection during a homologous flare. They concluded 
that the deflection was due to the open magnetic field at the edge of an AR, which developed into a coronal hole later. 

For the jet reported here, we attribute its apparent deflection to propagation along a curved pre-existing large-scale coronal
loop rather than any dynamic interaction with surrounding magnetic structures, as neither brightenings nor other signatures 
of interaction were observed at the site where the jet was deflected, as shown in the cases of \cite{zheng16} or 
\cite{li20}. Comparative analysis of the January 10 and 11 homologous jets suggests that an S-shaped trans-equatorial loop may 
span the western hemisphere, with its northern end anchored in AR 13536 and its southern end rooted in the 
weak, dispersed magnetic field region located to the south of AR 13543. The jet likely initiated through reconnection between 
the erupting F2's overlying arcade and this pre-existing loop system near its northern end, subsequently tracing the loop's 
bent shape to produce the observed $\sim 90^{\circ}$ apparent deflection, an effect enhanced by the visual distortion of 
projected effect. Moreover, the trans-equatorial loop may be located at a higher altitude than both quiescent filament 
QF and the main magnetic loop of AR 13543, thus the jet passed overhead without interacting with them.
    
In summary, this study provides a comprehensive analysis of a sympathetic filament eruption event and the associated jet 
that propagated along a curved path. The event began as a seemingly ordinary filament eruption, but subsequently 
led to a series of interconnected phenomena, including partial filament eruption, failed filament eruption, sympathetic filament 
eruption, jet produced by filament eruption, and apparent deflection of jet. Therefore, this event shows the 
complexity and diversity of solar activities and advances our understanding of the physical nature of solar eruptions.
 
\begin{acknowledgments}
We thank an anonymous referee for many constructive suggestions that improved the quality of this paper, and
the {\it SDO} and GONG teams for granting free access to their internet databases. This work uses the data 
from CHASE mission supported by China National Space Administration. This work is supported by the Strategic Priority Research 
Program of the Chinese Academy of Sciences, Grant No. XDB0560000, the National Key R\&D Program of China 2021YFA1600502,
the National Natural Science Foundation of China under grants 12273108, 12273106, 12350004, and 12173084, the ``Yunnan 
Revitalization Talent Support Program'' Innovation Team Project (202405AS350012), and by the Specialized Research Fund for 
State Key Laboratory of Solar Activity and Space Weather. H. C. was also supported by the Chinese Academy of Sciences (CAS)
Scholarship.
\end{acknowledgments}

\appendix
Following the partial eruption of F1, plasma was observed to flow along its spine toward the region beneath F2. 
We propose that this sustained mass transfer could increase the density and consequently the gas pressure below F2. To 
quantitatively assess the influence of this mass flow, we estimated the enhancement in gas pressure beneath F2 as follows:

Assuming that the mass flow from F1 to F2 possesses a cylindrical cross-section with a diameter $d_{1}$ equal to the flow's 
projected width, and that the flow speed $v$ is uniform and sustained over a time period $t$, the total volume of plasma 
transported to the region beneath F2 can be estimated as:
\begin{equation}
V_{1} = \pi {(\frac{d_{1}}{2})^{2}}vt
\end{equation}

Using AIA 304 \AA\ images, the projected width ($d_{1}$) of the plasma flow from F1 to F2 was measured as approximately 
$4^{''}.84$ through visual inspection, corresponding to $3.51 \times 10^{6}$ m on the solar surface. From time-slice plots 
constructed along slit AB, the mean velocity ($v$) of the flow was estimated to be $4.39 \times 10^{4}$ m s$^{-1}$, and the 
duration ($t$) from the flow's initial arrival at F2 to the subsequent brightening near F2 was determined to be $1770$ s. 
The resulting plasma volume ($V_{1}$) transported to F2 is therefore approximately $7.52 \times 10^{20}$ m$^{3}$.

Similarly, we assume that the injected material accumulated in a cylindrical volume beneath F2, with the cross-sectional 
diameter $d_{2}$ is comparable to the projected width of F2, and the height $h_{2}$ is comparable to the height of F2. 
The volume $V_{2}$ of the accumulated plasma can therefore be calculated as:
\begin{equation}
V_{2} = \pi {(\frac{d_{2}}{2})^{2}}h_{2}
\end{equation}
Here, $d_{2}$ was similarly determined from AIA 304 \AA\ images via visual inspection as $5^{''}.0$, equivalent to 
$3.63 \times 10^{6}$ m. The height of F2, $h_{2}$, was taken to be $2.0 \times 10^{6}$ m, consistent with typical AR 
filament heights which are generally below $10^{7}$ m \citep{filippov13}. Thus, the volume $V_{2}$ is calculated to be 
$2.07 \times 10^{19}$ m$^{3}$.

The mass of plasma accumulated beneath F2 must equal the total mass transported from F1 to F2, expressed as:
\begin{equation}
n_{e1}V_{1} = n_{e2}V_{2}
\end{equation}
Here, $n_{e1}$ denotes the electron density within the mass flow originating from F1, while $n_{e2}$ represents the 
electron density of the accumulated materia beneath F2, corresponding to the density increase resulting from the mass 
injection. According to \cite{labrosse10}, the typical electron density of filament plasma ranges from $10^{9}$ to 
$10^{11}$ cm$^{-3}$. Assuming $n_{e1} = 10^{10}$ cm$^{-3}$ (equivalent to $10^{16}$ m$^{-3}$ in the international 
system of units), $n_{e2}$ can be derived from the above equation as $3.63 \times 10^{17}$ m$^{-3}$.

The resulting increase in gas pressure below F2 can then be estimated using the ideal gas law:
\begin{equation}
P_{g} = n_{e2}kT
\end{equation}
where $k=1.38 \times 10^{-23}$ J K$^{-1}$ is the Boltzmann constant, and $T$ is the temperature of the injected plasma.
Although typical filament plasma temperatures range from $4300$ to $10000$ K \citep{labrosse10}, the injected material 
in this case was heated by F1's eruption. We therefore adopt $T=10^{5}$ K as a representative value, in line with
the reconnection-heated plasma temperature given by \cite{chen20a}. The gas pressure increase is estimated as approximately 
$0.5$ Pa, or $5.0$ dyn cm$^{-2}$.

On the other hand, the magnetic pressure of a filament can be estimated as
\begin{equation}
P_{m} = \frac{B^{2}}{8\pi}
\end{equation}
Assuming an average magnetic field strength of $B=20$ G for solar filaments \citep{mackay10}, the magnetic 
pressure of F2 is calculated to be approximately $15.92$ dyn cm$^{-2}$.

Nevertheless, it is important to emphasize that our estimation of the gas pressure is highly approximate. The 
assumption of cylindrical geometries for both the mass flow and the accumulated material is an idealization, and the diameters
were determined through visual inspection, which introduces subjectivity. Furthermore, key physical quantities including 
plasma density, temperature, and the height of filament exhibit considerable variations in the literature, each potentially
contributing substantial uncertainty to the result. While the estimation of magnetic pressure is comparatively more 
reliable, it remains subject to non-negligible uncertainties, primarily due to the challenges in accurately measuring 
filament magnetic field strengths.

\bibliography{ms}{}
\bibliographystyle{aasjournal}

\clearpage

\begin{figure}
\plotone{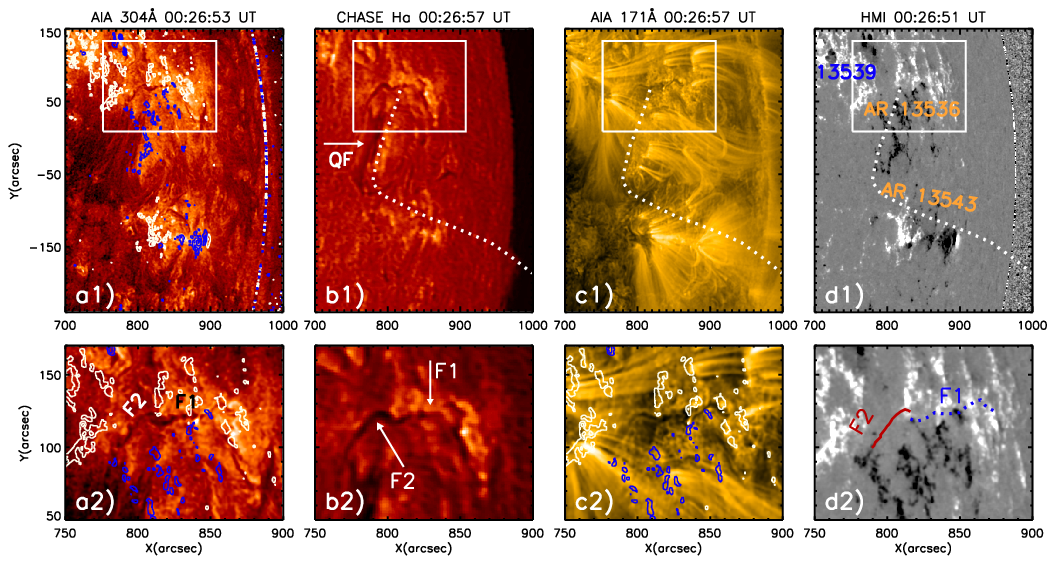}   
\caption{AIA 304 \AA\ (a1-a2), 171 \AA\ (c1-c2), CHASE H$\alpha$ (b1-b2) images and HMI LOS magnetograms (d1-d2) show the
filaments and their surroundings prior to eruption. The white rectangles in the top row show the field of view (FOV)
of the bottom row, while the white dotted curves indicate the trajectory of the following jet at 03:40:17 UT. 
The white arrows in panels (b1) and (b2) point to a quiescent filament QF and two filaments F1 and F2, while the blue and 
red curves in panel (d2) indicate the traced spines of F1 and F2 from panel (a2) by manual visual inspection. The 
white and blue contours superimposed on panels (a1), (a2) and (c2) represent the simultaneous HMI LOS magnetic field 
with intensities of $\pm 150$, $\pm 400$, and $\pm 600$ Gauss. 
\label{fig1}}
\end{figure}

\begin{figure}
\epsscale{0.95}
\plotone{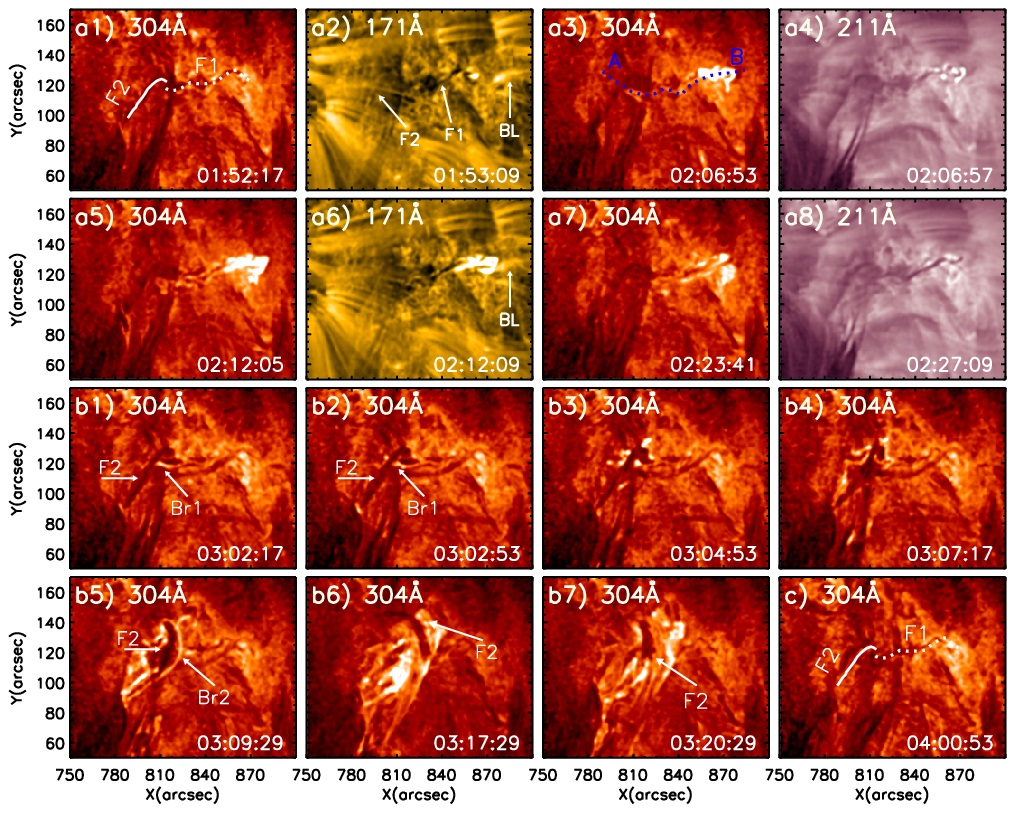}   
\caption{AIA 304 \AA, 171 \AA\ and 211 \AA\ images show the sympathetic eruption of filaments F1 and F2. The dotted
and solid white curves in panels (a1) and (c) indicate the spines of F1 and F2 obtained from 304 \AA\ image at 00:26:53 UT
by manual visual inspection, respectively. The dotted blue curve AB in panel (a3) shows the slit position of the 
time-slice plots in Figure 3. The white arrows in different panels point out some noteworthy features during the 
eruption. The FOV is the same as that of the bottom row of Figure 1. (An animation of AIA 304, 211, 171, and 335 \AA\ direct 
images with the same FOV accompanies this figure. The 29-second animation covers the period from 01:30 UT to 03:59 UT. 
The observation cadence of the images is 12 s.)
\label{fig2}}
\end{figure}

\begin{figure}
\epsscale{0.95}
\plotone{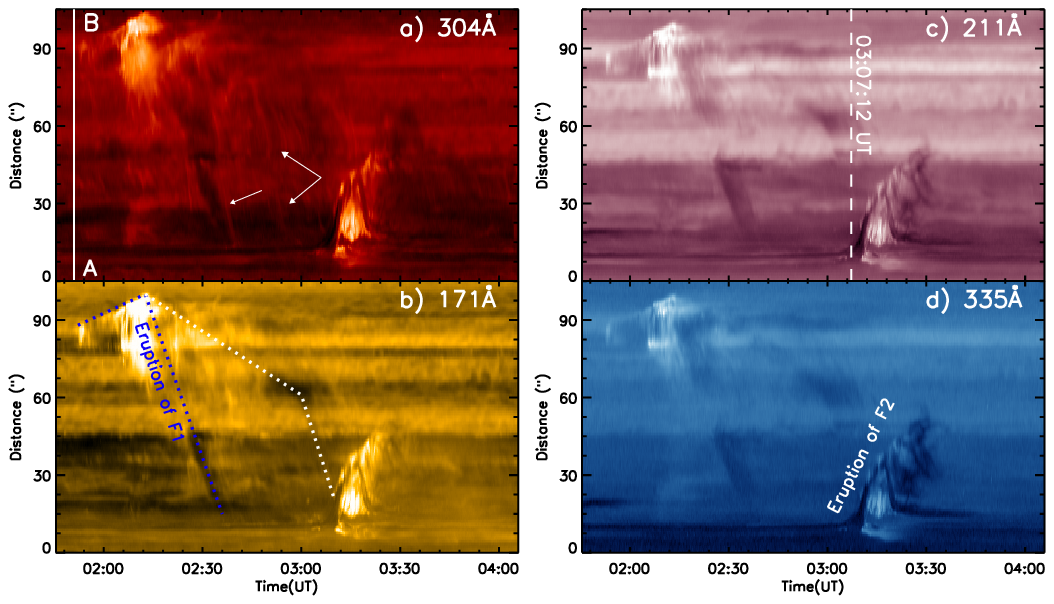}   
\caption{AIA 304, 171, 211, and 335 \AA\ time-distance plots. The time-slices are constructed along the slit AB shown in Figure 
2. The blue dotted lines in panel (b) mark the eruption of F1. The eruption and fall of F2 are visible in the right-hand
side of all panels. The white arrows in panel (a) and the white dotted lines in panel (b) point out some traces of 
plasma flow following the eruption of F1. The dashed white line indicates the approximate time of F2's eruption inferred from 
the time-slice of 211 \AA\ images.
\label{fig3}}
\end{figure}

\begin{figure}
\epsscale{0.95}
\plotone{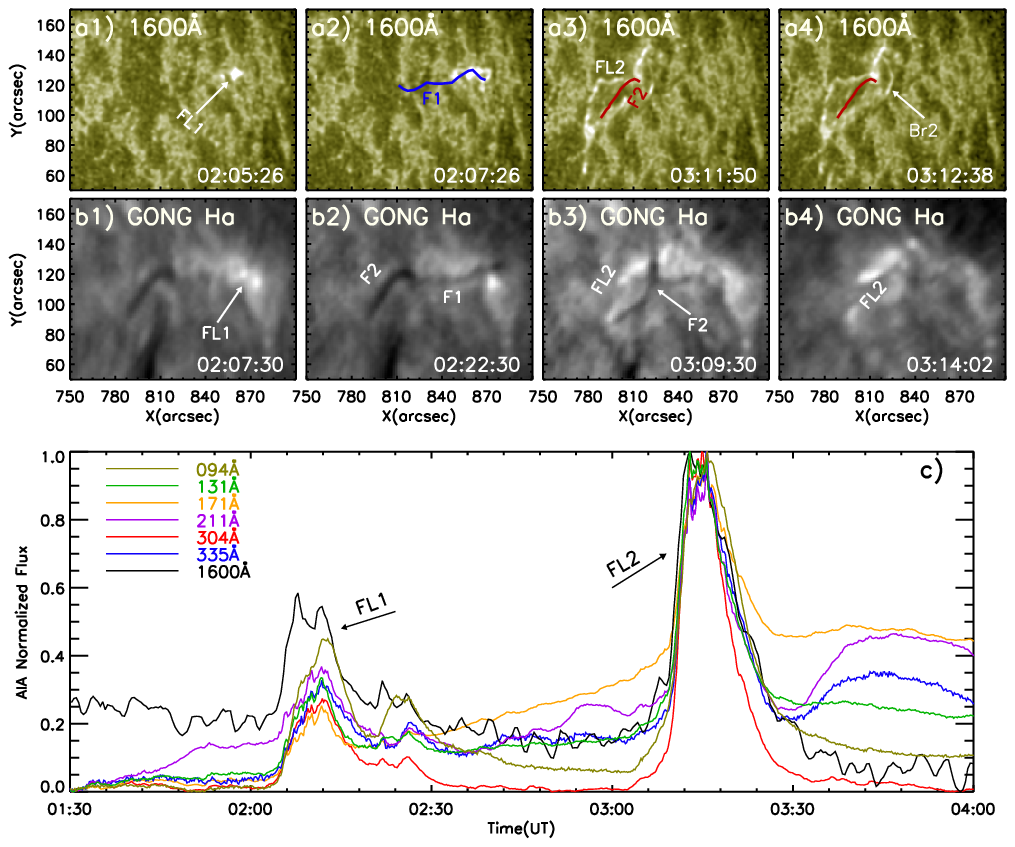}   
\caption{(a1-b4) Weak flares associated with eruptions of F1 and F2 showed in AIA 1600 \AA\ and GONG H$\alpha$
images. The axes of F1 and F2 obtained by manual visual inspection before the eruptions are superimposed as blue and 
red curves, respectively. The FOV is the same as in Figure 2. (c) Normalized light curves of the AIA 1600 \AA\ and six EUV 
wavebands calculated in the FOV of the above images.
\label{fig4}}
\end{figure}

\begin{figure}
\epsscale{0.9}
\plotone{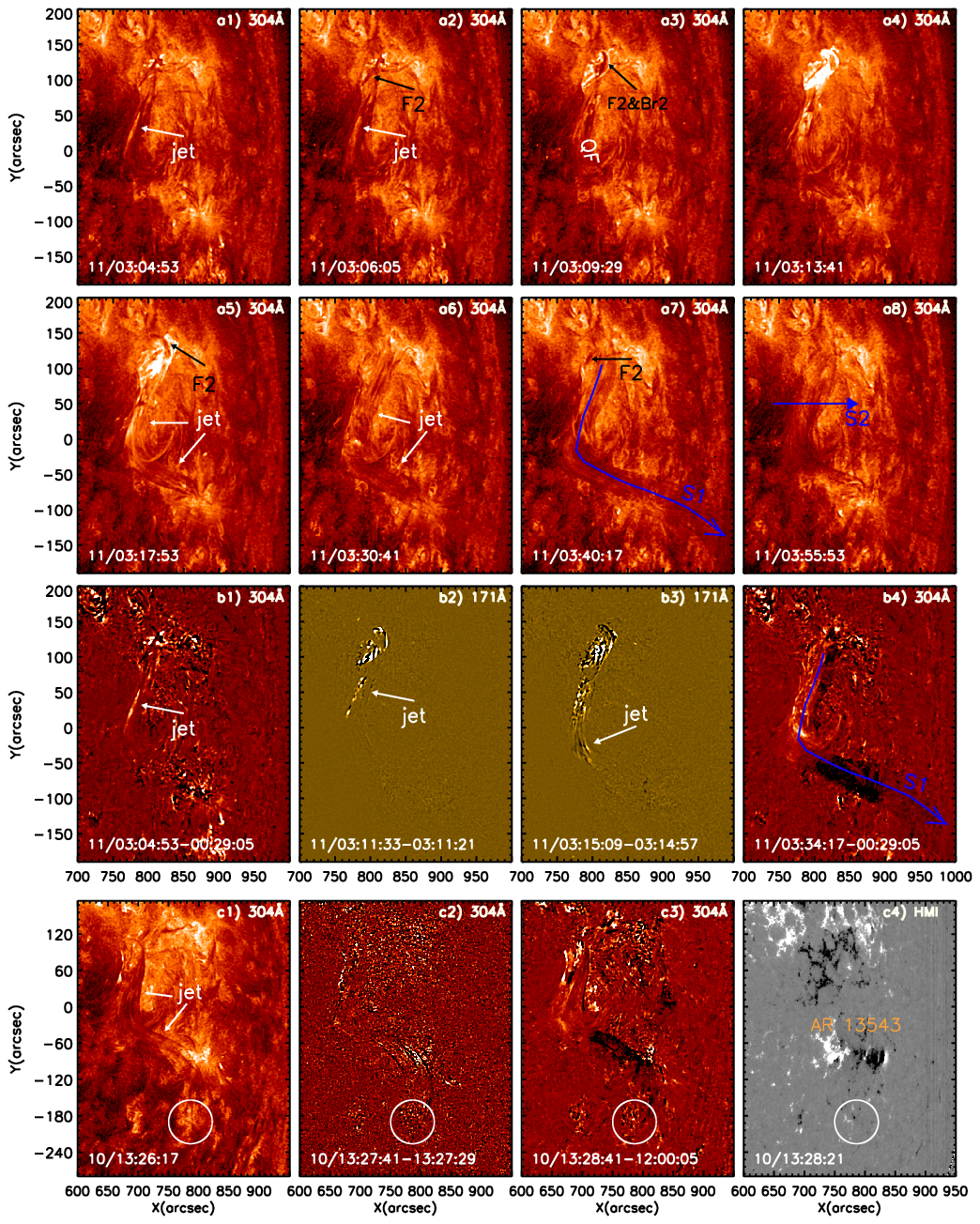}   
\caption{The coronal jet driven by the sympathetic filament eruptions. (a1-a8) AIA 304 \AA\ direct images. (b1-b4) AIA 
304 \AA\ fixed-base difference images and 171 \AA\ running difference images. The white arrows point out the jet, and 
the black arrows point out the erupting filament F2 and the brightening Br2. The blue arrows indicate the positions of two 
slits S1 and S2, which are used for the time-slice plots shown in Figure 6. The FOV is the same as that of the top row 
of Figure 1. (c1-c4) AIA 304 \AA\ images and HMI LOS magnetogram of the previous day showing a homologous jet occurred
in the same region. The white circles indicate an approximate region where the south end of the jet rooted.
(An animation of AIA 304 \AA\ direct images, 304 \AA\ fixed-base difference images, and 171 \AA\ running difference images
are available. This animation corresponds to the top three rows of the figure above, with the same FOV and a cadence of 12 s.
The animation lasts 29 seconds and covers the period from 01:30 UT to 03:59 UT on 2024 January 11. ) 
\label{fig5}}
\end{figure}

\begin{figure}
\epsscale{0.7}
\plotone{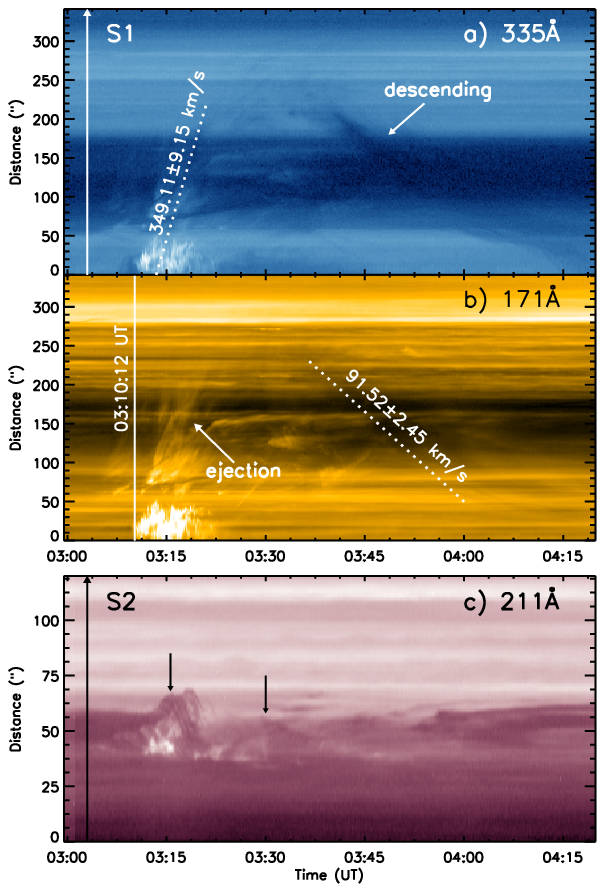}   
\caption{(a-b) Time-distance plots along the jet spire in AIA 335 and 171 \AA. (c) Time-space plot along a slit perpendicular 
to the jet spire in AIA 211 \AA. The vertical white line indicates the start time of the jet. The two short white arrows
point out the ejection and the descending phases of the jet respectively, and the dotted lines depict the linear fitting of
the trajectory of these phases. The short black arrows show two episodes of the lateral motion of the jet. The positions 
of two slits S1 and S2 are indicated by blue arrows in Figure 5.
\label{fig6}}
\end{figure}

\begin{figure}
\epsscale{0.95}
\plotone{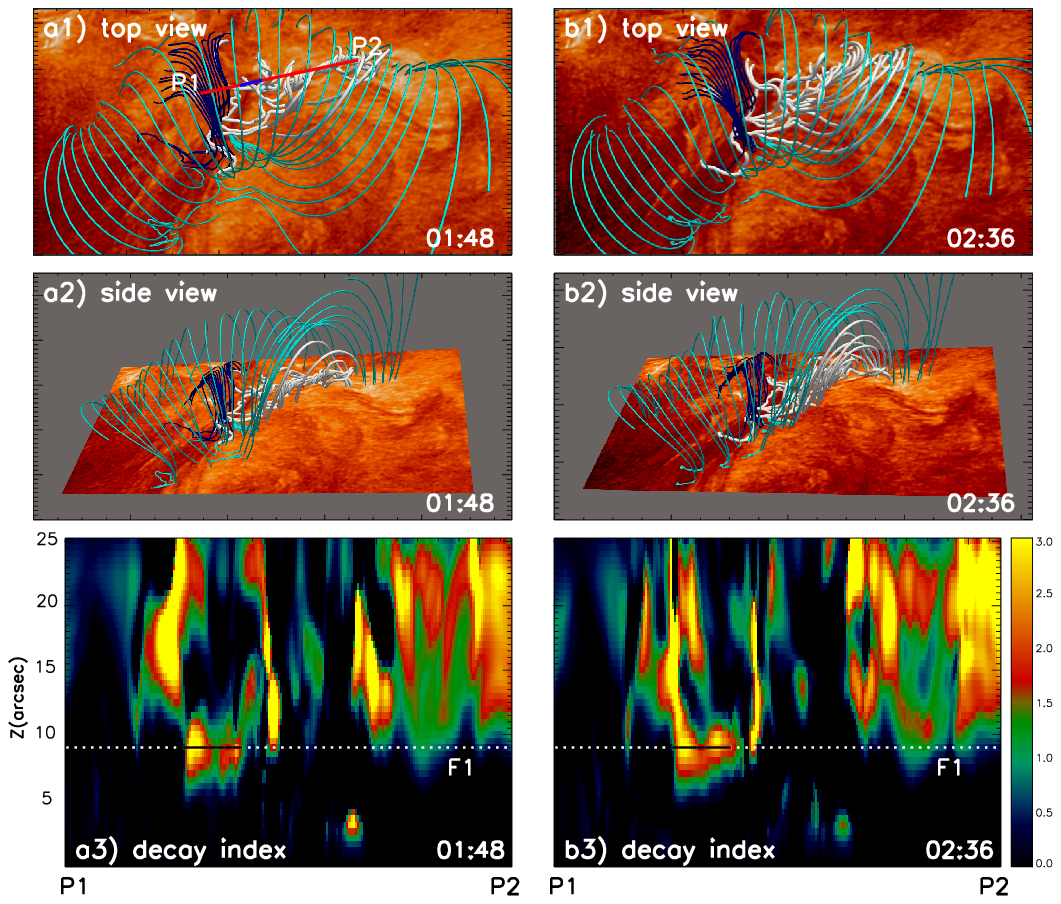}   
\caption{NLFFF extrapolation of the coronal magnetic topology and the decay index above the two filaments before 
((a1)-(a3)) and after ((b1)-(b3)) F1's eruption. The background image is the AIA 304 \AA\ image captured at 01:44:53 UT. 
The overlying curves represent the extrapolated coronal magnetic field lines, among which the grey curves indicate the flux 
ropes of F1, the dark blue curves indicate the magnetic arcade above F2, and the cyan curves represent a common arcade 
covering both F1 and F2. The horizontal axes of panels (a3) and (b3) correspond to the spatial coordinate along 
the red line P1P2 in panel (a1). The white dotted lines in the bottom row indicate the height of the flux rope. The
overlaid black lines on these lines correspond to the region shown by the blue segment of P1P2 in panel (a1).}
\label{fig7}
\end{figure}
\end{document}